# The evolving absolute magnitude of type 1a supernovae and its critical impact on the cosmological parameters.

A.P. Mahtessian[1], G.S. Karapetian, M. A. Hovhannisyan[2], L.A. Mahtessian[2]

[1]Byurakan Astrophysical Observatory (BAO) after V. Ambartsumian NAS of the Republic of Armenia
Byurakan, Aragatzotn Province, Republic of Armenia, 0213
[2] Institute of Applied Problems of Physics NAS of the Republic of Armenia 25 Hrachya Nersissian Str.,
Yerevan, Republic of Armenia, 0014

**Abstract.** In this work, a computer optimization model has been developed that allows one to load the initial observation data of supernovae 1a into a table and, simply, by searching for the best fit between observations and theory, obtain the values of the parameters of cosmological models. Naturally, the initial data are redshifts z and apparent magnitudes m at the maximum brightness of supernovae.

For better fit between theory and observation, Pearson's Chi$^2$ (Chi-squared) goodness-of-fit test was used. The results are obtained for the ΛCDM model and, for comparison, the model with a zero cosmological constant.

In order to improve the fit between observed data and theory, the optimization is carried out assuming that the absolute magnitude of supernovae is not constant, but evolves with time. It is assumed that the dependence of the absolute magnitude on the redshift is linear: $M=M(z=0)+\varepsilon_c z$, where $\varepsilon_c$ is the evolution coefficient of the absolute magnitude of type 1a supernovae.

In the case of a flat universe ($\Omega_M+\Omega_\Lambda=1$), the best fit between theory and observation is $\varepsilon_c=0.304$. In this case, for the cosmological parameters we obtain $\Omega_\Lambda=0.000$, $\Omega_M=1.000$. And for the absolute magnitude of supernovae 1a, we obtain the value -18.875. Naturally, this result exactly coincides with the simulation result for the model with a zero cosmological constant ($\varepsilon_c=0.304$, $q_0=0.500$, $M_0=-18.875$).

Within the framework of the Friedmann-Robertson-Walker model, without restriction on space curvature ($\Omega_M+\Omega_\Lambda+\Omega_K=1$), we obtain the following values: $\varepsilon_c=0.304$, $\Omega_\Lambda=0$, $\Omega_M=1.000$, $\Omega_K=0.000$, $M_0=-18.875$. Those, the general case also leads to a flat Universe model ($\Omega_K=0.000$).

Within the framework of this work, the critical influence of the absolute magnitude M of type 1a supernovae on the cosmological parameters is also shown. In particular, it was found that a change in this value by only 0.4$^m$ (from -19.11 to -18.71) leads to a change in the parameters from $\Omega_\Lambda=0.7$ and $\Omega_M=0.3$ to $\Omega_\Lambda=0$ and $\Omega_M=1$. We note that the distribution of the absolute magnitudes of supernovae 1a has a rather large width, which leads us to the idea that one must be extremely careful when determining the value of the absolute magnitude of supernovae. Determining this value from several well-studied stars can lead to erroneous values for the cosmological parameters. For this, it is proposed that the determination of the value of the absolute magnitude of supernovae should also be subject to simulation. In this paper, the average absolute magnitude of supernovae is determined by all the stars in the sample.

This is the main reason why our results disagree with the results of other authors in our previous article, where the case with a constant absolute magnitude of type 1a supernovae was investigated.

The validity of our results is substantiated by the absolute magnitude test we proposed in the previous article.

Keywords: Supernovae SNe1a – Absolute Magnitude Evolution– Cosmological Parameters – Cosmology – Acceleration

1. Introduction

It is believed that a type Ia supernova is formed when a white dwarf captures matter from its neighbor in a binary system, as a result of which its mass increases to a possible limit - the Chandrasekhar limit, when already degraded electrons cannot resist gravitational pressure and the star passes into an unstable stage. An increase in the temperature and density of the star makes it possible for carbon and oxygen to be converted into $^{56}$Ni, which is accompanied by a thermonuclear explosion (Fowler and Hoyle 1960). The brightness of the star increases so much that sometimes it exceeds the brightness of the host galaxy, and it can be seen for several thousand megaparsecs. The mass of the exploded star is always near the Chandrasekhar limit, so in the case of such explosions the absolute magnitude can only vary within small limits. This allows these stars to be used as distance indicators (Sandage and Tammann 1982).

This feature of type Ia supernovae makes it possible to study the behavior of the Universe at considerable distances and evaluate the validity of one or another cosmological model.

At the beginning of the 20th century, Hubble obtained a very interesting result, which led to the conclusion that the universe is expanding, and the expansion rate is directly proportional to the distance from the observer. Hubble's work is based on the fact discovered by Slipher, that the spectral lines in the spectra of galaxies are shifted towards the long wavelength (Slipher 1924). Hubble found that this shift increases with increasing distances to galaxies (Hubble, 1929).

Another method for determining distance is based on the modulus of distance.

$$M = m - 5 log M_L - 25, \qquad (1)$$

where m is the apparent magnitude, M is the absolute magnitude, $D_L$ is the Luminosity distance.

When calculating the distance using this method, it is necessary to accurately estimate the value of the apparent magnitude of the object (take into account the galactic extinction, K-correction, spectral region, etc.). The absolute magnitude should be known either from theoretical approaches (for example, for type Ia supernova stars) or from empirical relationships (for example, in the case of Cepheids).

Riess et al. (1998) and Perlmutter et al. (1999), in order to study the properties of the universe, made two assumptions:

a. Assume that Ia-type supernovae are indicators of distances, that is, their absolute magnitudes can be considered constant.

b. That the Friedmann-Robertson-Walker (FRW) cosmological model, for the case of a flat universe, accurately describes the Universe.

Quite accurately taking into account the phenomena that can influence the result, they calculated the apparent magnitudes, and compared them with the values obtained from the cosmological model. It turned out that the apparent brightnesses were weaker than those obtained from the theory, that is, these objects are further away than they would be, based on Hubble's law. This led to the idea that the universe is expanding at an accelerating rate. In this regard, the idea of "dark energy" was introduced.

To avoid the idea of dark energy, various attempts have been made to explain the discrepancy between the theoretical and observed supernova luminosities by other phenomena. Let's list some of them.

a. The weakening of the apparent magnitude of a star occurs due to the absorption or scattering of light by matter in the path of light.

b. There is an evolution in the luminosity of a white dwarf, depending on the chemical composition of the host galaxy over time.

c. Gravity lenses.

d. The reason is the uneven distribution of matter in the Universe.

e. It is assumed that there are two types of supernovae Ia in nature. The second type is not numerous and is formed from the merger of two white dwarfs. As a result of the merger, the mass of the exploding star is no longer fixed.

f. Observational errors may also increase due to the fact that the brightness curves of various supernovae are recorded under different conditions (on Earth and in space)

The degree of influence of these phenomena has been discussed in various studies, showing that many of these inaccuracies cannot be considered satisfactory for refuting the results obtained by Riess et al. (1998) and Perlmutter et al. (1999). They can be found in Weinberg (2008).

However, in our opinion, there are two observational facts that cannot be ignored.

**First.** This is the rather large width of the distribution of the absolute magnitudes of type Ia supernovae. This issue was studied in the article by Ashall et al. (2016). The average absolute magnitude of 115 studied stars was obtained <MB> = -19.04 ± 0.07, standard deviation σMB = 0.70. 89 of them have late host galaxies (Sa-Irr or star-forming galaxies, S-F), for which <MB> = -19.20 ± 0.05, σMB = 0.49, and 26 have early host galaxies (E-SO or passive galaxies), respectively <MB> = -18.48 ± 0.19, σMB = 0.98.

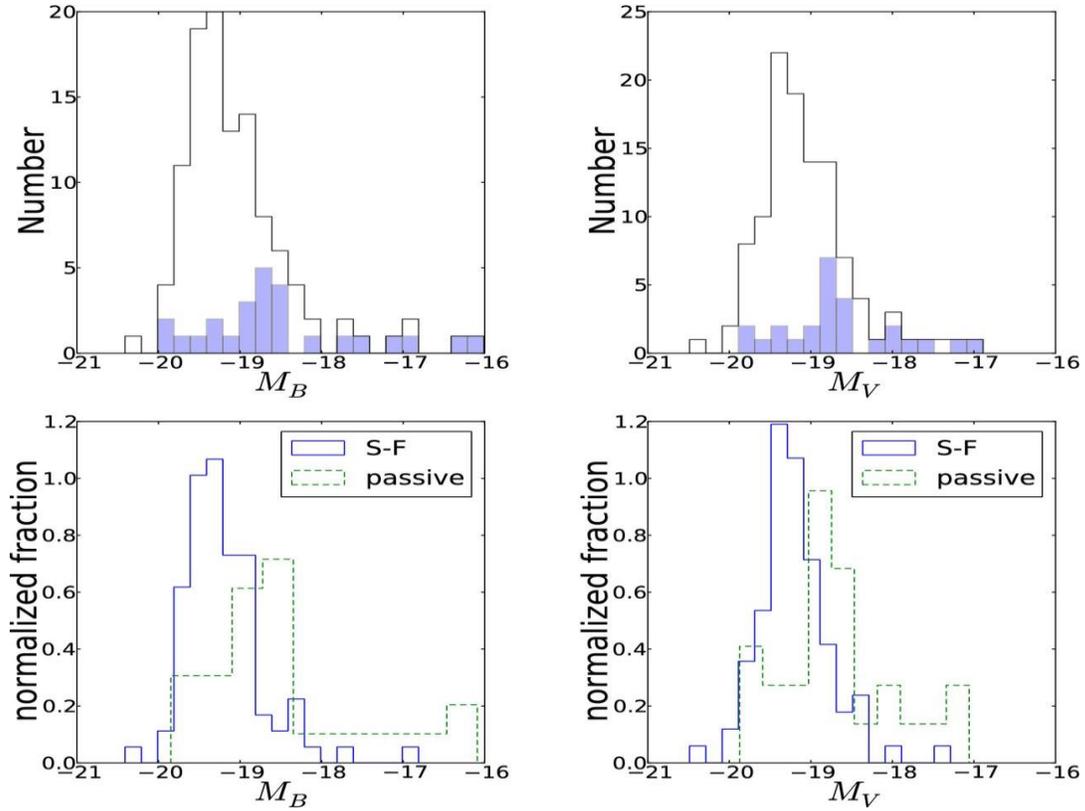

Fig.1. Distribution of absolute magnitudes of 115 type 1a supernovae. Graph copied from Ashall et al. (2016)

Such large standard deviations in the absolute magnitude distributions of type Ia supernovae allow us to conclude that when estimating the values of cosmological parameters, it is wrong to take as a basis the absolute magnitude determined by few stars. Mahtessian et al. (2020) (Paper I) showed that in this case the obtained cosmological parameters lead to a violation of the initial assumption that the absolute magnitudes of type Ia supernovae do not change with distance. This violation disappears when the absolute magnitude of supernovae is estimated while estimating the cosmological parameters.

Thus, when estimating cosmological parameters, the absolute magnitude of supernovae should also be an estimated parameter. The absence of such an approach can be considered a shortcoming in the works of other authors related to this topic.

Note that this approach also improves the fit between the observational data and the theory. Assuming that the absolute magnitude of supernovae is constant with distance, we get that the share of dark energy in a flat universe does not exceed 50%. In Mahtessian et al. (2020) also obtained another important result that the cosmological model with a zero cosmological parameter describes the universe no worse than the Friedmann-Robertson-Walker model.

**Second.** The correlation between the absolute magnitude of supernovae and the age of the stellar population of host galaxies indicates that there is an evolution in the absolute magnitude of supernovae (Kang et al. 2020).

It is known that the absolute magnitude of type 1a supernovae correlates with the characteristics of the host galaxy. For example, in Hicken et al. (2009b) found a systematic difference in the absolute magnitude of supernovae of ~0.14 magnitude between very early and very late galaxies. Sullivan et al. (2010) and Kelly et al. (2010) found that SNe Ia in less massive galaxies (by a factor of 10) are weaker by ~0.08 magnitudes than in more massive galaxies. Rigault et al. (2018) showed that SNe Ia in environments with local star formation (higher local SFR) is about 0.16 magnitudes weaker than in locally passive environments (lower local SFR).

Kang et. al. (2020), noted features of the host galaxies (morphology, mass and local SFR) were converted to age differences with methods known in the literature. Table 1 is taken from Kang et al.

(2020). The table shows the correlation of the absolute magnitude of supernovae 1a with the properties of the parent galaxies. The last column of Table 1 shows the estimated absolute magnitude evolution over 5.3 Gyr, which corresponds to the difference in age at z=0 and z=1 (see Kang et. al. 2020), for each of the four different studies. The average of these values is ~0.25 mag/5.3 Gyr. In this range of redshifts, the observed decrease in supernova brightness in the Hubble diagram is approximately comparable to this value (see, for example, Riess et al. 1998). And so, this effect may be associated with the evolution of the luminosity of supernovae and has nothing to do with the accelerated expansion of the universe.

Table 1. Correlation of the absolute magnitude of supernovae 1a with the properties of host galaxies Kang et. al. (2020).

| Host Property | Reference | Original Correlation | Direction | Converted to Age difference |
|---|---|---|---|---|
| Morphology | Hicken et al. (2009) | ΔHR/Δmorph. ≈0.14 mag/(Scd/Irr ~E/S0) | Fainter in Later type galaxy | 0.19 mag/5.3 Gyr Fainter in Younger galaxy |
| Mass | Sullivan et al. (2010) | ΔHR/Δmass ≈0.08 mag/(Δlog M⋆~1 ) | Fainter in Less massive galaxy | 0.21 mag/5.3 Gyr Fainter in Younger galaxy |
| Local SFR | Rigault et al. (2018) | ΔHR/Δlocal SFR ≈0.16 mag/(Δlog LsSFRstep ~ 2 yr~1 kpc−2) | Fainter in Higher SFR environments | 0.34 mag/5.3 Gyr Fainter in Younger galaxy |
| Population Age | Kang J., et. al. 2020 | ΔHR/Δage ≈0.051 mag/Gyr (YEPS) | Fainter in Younger galaxy | 0.27 mag/5.3 Gyr Fainter in Younger galaxy |

Thus, when estimating the cosmological parameters, it is also important to estimate the possible evolution of the absolute magnitudes of supernovae.

In this article, we study models of the universe under the assumption of the existence of an evolution of the luminosities of type 1a supernovae and try to find those values of the cosmological parameters for which there will be the best fit between theory and observation.

2. Theory.

In this paper, we will discuss two models: the ΛCDM model used by Riess et al. (1998) and Perlmutter et al. (1999) for the case of a flat universe, and the model with a zero cosmological constant, which was widely used before these works (until 1999). The first model assumes the existence of dark energy; in the second model, such a hypothesis is not necessary.

In the case of the ΛCDM model, the dependence of the luminosity distance on redshift is given by the following formula:

$$D_L(z, \Omega_M, \Omega_\Lambda, \Omega_K) = CH_0^{-1}(1+z)|\Omega_K|^{-1/2}$$
$$\times \sinn\{|\Omega_K|^{1/2} \int_0^z dz\, [(1+z)^2(1+\Omega_M z) - z(2+z)\Omega_\Lambda]^{-1/2}\} \qquad (2)$$

where z is the redshift of the object. $H_0$ is the Hubble constant. $\Omega_K$ is related to the curvature of space and in the case of a flat space it is equal to 0 (Carroll et al., 1992): $\Omega_K = 1 - \Omega_M - \Omega_\Lambda$, sinn = sinh when $\Omega_K \geq 0$ and sinn = sin when $\Omega_K \leq 0$: in the case of $\Omega_K = 0$ we will have:

$$D_L(z, \Omega_M, \Omega_\Lambda) = \frac{C(1+z)}{H_0} \int_0^z dz\, [(1+z)^2(1+\Omega_M z) - z(2+z)\Omega_\Lambda]^{-1/2} \qquad (3)$$

or

$$D_L(z, \Omega_M, \Omega_\Lambda) = \frac{C(1+z)}{H_0} \int_0^z dz \left[(1+z)^3 \Omega_M + \Omega_\Lambda\right]^{-1/2}$$

If we assume that $\Omega_\Lambda=1$ and $\Omega_M=0$, we will have (Weinberg, 2008)

$$D_L(z) = \frac{C}{H_0}(z + z^2) \tag{4}$$

If $\Omega_\Lambda = 0$, a $\Omega_M = 1$, will have

$$D_L(z) = \frac{2C}{H_0}\left[(1+z) - \sqrt{1+z}\right] \tag{5}$$

It should be noted that in 1998, prior to the work of Riess et al. (1998) and Perlmutter et al. (1999) the equations of general relativity (GR) were commonly used with a zero cosmological constant ($\Omega_\Lambda=0$). Using this model, Mattig (1958) integrated these equations exactly and obtained the luminosity distance as a function of redshift.

$$D_L(z, q_0) = \frac{C}{H_0 q_0^2}\left[q_0 z + (q_0 - 1)(\sqrt{1 + 2q_0 z} - 1)\right] \tag{6}$$

where $q_0$ is the deceleration parameter, in this case:

$$q_0 = \frac{\Omega_M}{2}$$

(6) at $q_0=0.5$ coincides with (5).

For the luminosity distance in the case of a flat universe we will use formula (3), for the luminosity distance in the model with zero cosmological constant ($\Omega_\Lambda=0$) we will use formula (6).
We will also discuss the general case (2) with nonzero space curvature. We also assume that the dependence of the absolute magnitude of the supernova on z is linear. Then we can assume $M_z=M_0+\varepsilon_c z$ and formula (1) can be written as follows:

$$M_0 + \varepsilon_c z = m - 5\lg D_L - 25, \tag{7}$$

Where $M_0$ is the absolute magnitude of the supernova at z=0, $\varepsilon_c$ is the evolution coefficient of the absolute magnitude.

3. Procedure

Our approach was as follows: to develop a computer model where one can load the observed data of supernovae 1a and easily obtain cosmological parameters by achieving the best fit of observations with theory.
As the search variables of the computer model, both the cosmological parameters of the Friedmann-Robertson-Walker model $\Omega_\Lambda$, $\Omega_M$, $\Omega_K$ and the parameters $M_0$ and $\varepsilon_c$ were used.
The value of the Hubble constant $H_0$ is assumed to be 72.305 km/s/Mpc.
Solutions were sought in the ranges of variables:
$0 \leq \Omega_M \leq 1$,
$0 \leq \Omega_\Lambda \leq 1$,
$0 \leq \Omega_K \leq 1$,
$-19.5 \leq M_0 \leq -18$,
$-1 \leq \varepsilon_c \leq 1$,
under the condition of a flat universe ($\Omega_M+\Omega_\Lambda=1$) and without it ($\Omega_M+\Omega_\Lambda+\Omega_K=1$).
A model with a zero cosmological constant is also investigated, in which the search parameters are $q_0$, $M_0$ and $\varepsilon_c$. The solution was searched in the ranges:
$0 \leq q_0 \leq 0.5$,

-19.5 ≤ $M_0$ ≤ -18,
-1 ≤ $\varepsilon_c$ ≤ 1.
Solver Excell, Macroses and SciDAVIs were used as optimization decision tools.

4. Sample

For the study, we use a subsample from SNe1a "Union2" (Amanullah et al. 2010; http://vizier.u-strasbg.fr/viz-bin/VizieR-3?-source=J/ApJ/716/712). The sample consists of 719 supernovae identified in 17 papers (Hamuy et al. 1996; Krisciunas et al. 2005; Riess et al. 1999; Jha et al. 2006; Kowalski et al. 2008; Hicken et al. 2009a, 2009b; Holtzman et al. 2008 Riess et al 1998 Perlmutter et al 1999 Barris et al 2004 Amanullah et al 2008 Knop et al 2003 Astier et al 2006 Miknaitis et al 2007 Tonry et al. 2003; Riess et al. 2007; Amanullah et al. 2010). Following several principles, the authors (Amanullah et al. 2010) cleared the sample and retained 557 supernovae for further study. We will also use the observational material of these 557 stars without making any changes.

5. Result

We analyze the Hubble diagram and find those values of the parameters present in the discussed model of the universe, which provide the best fit between the model and observation.
On the Hubble diagram, the theoretical curve can be represented by the following relationship:

$$B_{mag}^{th}(z, \Omega_M, \Omega_\Lambda, \Omega_K) = M_0 + \varepsilon_c z + 5 \times \log D_L(z, \Omega_M, \Omega_\Lambda, \Omega_K) + 25 \qquad (8)$$

for the Friedmann-Robertson-Walker model, or

$$B_{mag}^{th}(z, q_0) = M_0 + \varepsilon_c z + 5 \times \log D_L(z, q_0) + 25 \qquad (9)$$

for the model with a zero cosmological constant.
We need to find those values of the parameters $\Omega_M, \Omega_\Lambda, \Omega_K, M_0, \varepsilon_c$ in the first case and $q_0, M_0, \varepsilon_c$ in the second case, so that the sum of squares $(B_{obs} - B_{mag}^{th}(z))$ would be minimal:

$$Chi^2 = \sum (B_{obs} - B_{mag}^{th}(z))^2 = \min$$

For the luminosity distance $D_L(z)$ we use formulas (2) and (3) for the Friedmann-Robertson-Walker model and formula (6) for the model with zero cosmological constant.
We will investigate the following cases:

A. The flat universe of Friedmann-Robertson-Walker ($\Omega_\Lambda + \Omega_M = 1$). We will look for $\Omega_\Lambda$, $\Omega_M$ and $M_0$, $\varepsilon_c$ for the best fit between theory and observation. We will discuss two cases:
a - there is no evolution of the absolute magnitudes of supernovae ($\varepsilon_c = 0$) and
b - there is an evolution of the absolute magnitudes of supernovae ($\varepsilon_c \neq 0$).

B. The Friedmann-Robertson-Walker universe without a restriction on space curvature ($\Omega_\Lambda + \Omega_M + \Omega_K = 1$). We will look for $\Omega_\Lambda$, $\Omega_M$, $\Omega_K$, and $M_0$ for the best fit between theory and observation. We will also discuss two cases:
a - $\varepsilon_c = 0$ and
b - $\varepsilon_c \neq 0$.

C. The universe with zero cosmological constant. We will look for $q_0$ and $M_0$ for the best fit between theory and observation. We will also discuss two cases:
a - $\varepsilon_c = 0$ and
b - $\varepsilon_c \neq 0$.

All cases are tested by the absolute magnitude test we proposed in Paper I.

5.1. The case $\Omega_K = 0$, $\Omega_\Lambda + \Omega_M = 1$, $\varepsilon_c = 0$, where $M_0$, $\Omega_\Lambda$, $\Omega_M$ are evaluated. Comparison with Amanullah et al. (2010)

In Table. 2, the case of a flat universe is considered without the assumption of the evolution of the absolute magnitude of type 1a supernovae. Here and in the tables that follow, the first column lists the parameters discussed, the second column is "Yes" when the parameter is evaluated in simulation, or "No" when the parameter is assumed to be constant. The third column shows the range of values for the parameter you are looking for. The fourth column shows the value of the desired parameter, at which the best fit between the observational data and the theory is observed (i.e., the minimum value of $Chi^2$ is obtained - the sum of the squared deviations of the observation points from the theoretical curve (Pearson's criteria) on the Hubble diagram). The fifth column shows the minimum value of $Chi^2$.

Table 2 shows that, assuming a constant absolute magnitude of supernovae 1a, we get $\Omega_\Lambda=0.4$, $\Omega_M=0.6$. This result is consistent with the result of Paper I (Mahtessian et al., 2020), where only the case with a constant absolute magnitude of type 1a supernovae was studied. It was noted above that we use the same sample with the same observational data as used in Amanullah et al. (2010). But these authors obtained $\Omega_\Lambda=0.73$, $\Omega_M=0.27$.

How can such a large difference be explained? At the top, we noted that with a large spread in the absolute magnitudes of supernovae 1a, in our opinion, it is not correct to use the average absolute magnitude obtained by several stars, and it is necessary that the absolute magnitude be obtained by the simulation method. Thus, with this approach, according to the studied sample, assuming the constancy of the absolute magnitude of supernovae, the fraction of dark energy is 0.4. Paper I studied different sub-samples of the Union (Kowalski et al. 2008) and Union 2 (Amanullah et al. 2010) compilations of type 1a supernovae. It has been found that, assuming a constant absolute magnitude of SNe1a, the fraction of dark energy does not exceed 0.5.

Table 2. The result of the search for the values of the parameters $M_0$, $\Omega_\Lambda$, $\Omega_M$ for the Flat Universe ($\Omega_\Lambda+\Omega_M=1$, $\Omega_K=0$) without the assumption of the evolution of the absolute magnitude SNe1a.

| Parameter | Variable | Search range | Parameter search result | $Chi^2$ |
|---|---|---|---|---|
| $M_0$ | Yes | -19.5 ÷ -18.0 | -18.903 | 83.7439 |
| $\varepsilon_c$ | No | 0 | 0 | |
| $\Omega_\Lambda$ | Yes | 0 ÷ 1 | 0.397 | |
| $\Omega_M$ | Yes | 0 ÷ 1 | 0.603 | |
| $\Omega_K$ | No | 0 | 0 | |

Now let's check the validity of our result with the test of the absolute magnitude proposed by us in Paper I. The meaning of the test is that after finding the values of the cosmological parameters, the dependence of the absolute magnitudes of SNe1a on the distance (on the redshift z) is plotted and its compliance with the initial assumption is checked.

Figure 2 plots the absolute magnitudes of SNe1a calculated from the parameters $\Omega_\Lambda=0.397$ and $\Omega_M=0.603$ depending on the redshift.

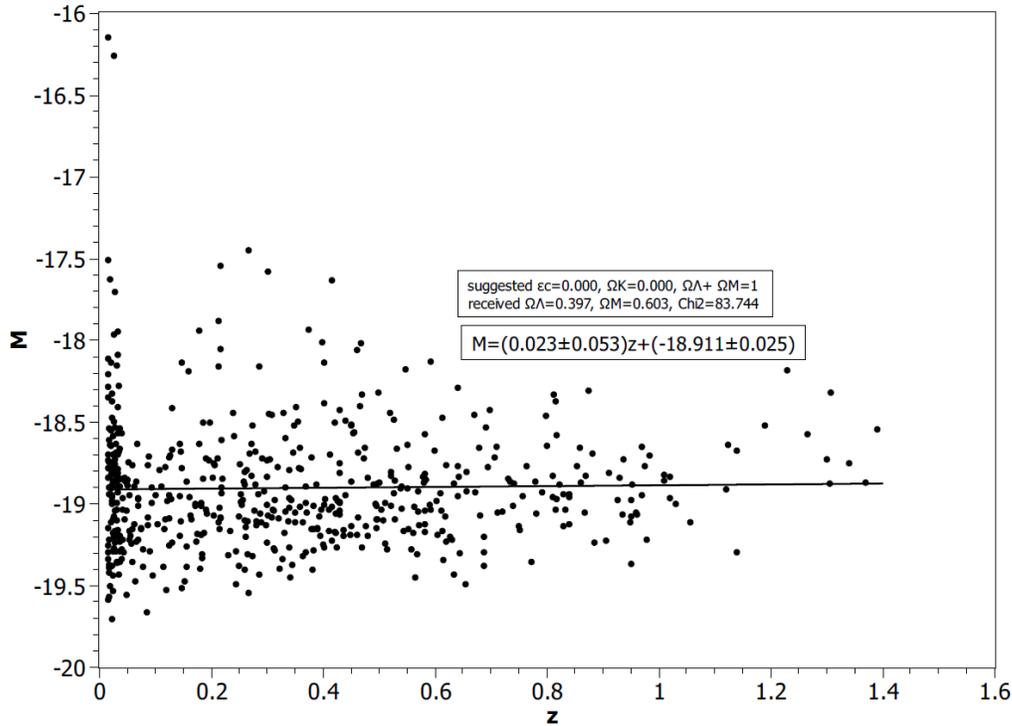

Fig.2. The absolute magnitudes of SNe1a calculated from the parameters $\Omega_\Lambda=0.397$ and $\Omega_M=0.603$ depending on the redshift.

As can be seen from Fig. 2, there is no noticeable relationship between M and z, i.e., the original assumption about the independence of the absolute magnitude of the redshift is observed.

In Paper I (where the case of distance-independent absolute magnitude of supernovae was investigated), different sub-samples from the Union (Kowalski, M. et al. 2008) and Union 2 (Amanullah et al. 2010) compilations were studied, and in all cases, after simulation, we reach the original assumption about the independence of the absolute magnitude of supernovae from redshift.

Under the assumption that the absolute magnitudes of supernova are constant, Amanullah et al. (2010) obtained the value $\Omega_\Lambda=0.73$ and $\Omega_M=0.27$. Let's test the absolute magnitude. The dependence of the absolute magnitude of SNe1a on the redshift at $\Omega_\Lambda=0.73$, $\Omega_M=0.27$ is shown in Fig. 3 (see also Paper I).

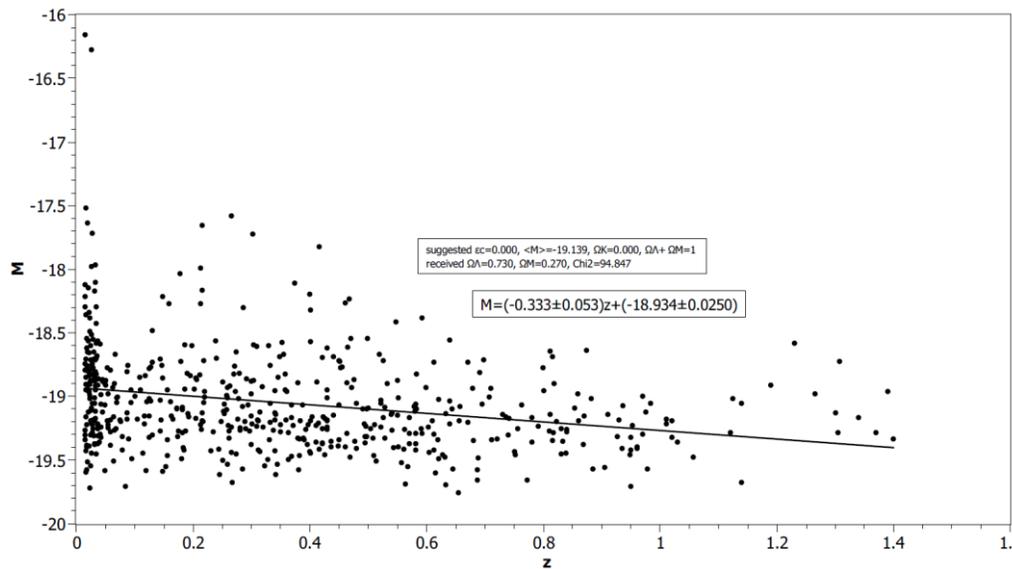

Fig. 3. Dependence of the absolute magnitude of SNe1a on the redshift at $\Omega_\Lambda=0.73$, $\Omega_M=0.27$.

As can be seen from Fig. 3 there is a clear relationship between the values under consideration. Thus, in this case, the assumption that the absolute magnitudes of SNe1a are independent of the redshift is violated. This gives grounds to believe that the authors found incorrect values of $\Omega_\Lambda$ and $\Omega_M$. This is also confirmed by the Chi$^2$ values. Their Chi$^2$ value is 94.85 (see Fig. 3), while ours is 83.74 (Table 2, Fig. 2).

It is noteworthy that the average absolute magnitude of SNe1a $M_0$ in Figures 2 and 3 differ little. This means that the obtained values of the cosmological parameters depend very strongly on the previously accepted average absolute magnitude of supernovae. This issue will be explored in detail below.

5.2. Case $\Omega_K=0$, $\Omega_\Lambda+\Omega_M=1$, estimated $M_0$, $\varepsilon_c$, $\Omega_\Lambda$, $\Omega_M$

In table. 3, the case of a flat universe is considered with the assumption of the evolution of the absolute magnitude of type 1a supernovae.

Table 3. The result of the search for the values of the parameters $M_0$, $\Omega_\Lambda$, $\Omega_M$ and $\varepsilon_c$ for the Flat Universe ($\Omega_\Lambda+\Omega_M=1$, $\Omega_K=0$) with the assumption of the evolution of the absolute magnitudes SNe1a.

| Parameter | Variable | Search range | Parameter search result | Chi$^2$ |
|---|---|---|---|---|
| $M_0$ | Yes | -19.5 ÷ -18.0 | -18.875 | |
| $\varepsilon_c$ | Yes | -1 ÷ 1 | 0.304 | |
| $\Omega_\Lambda$ | Yes | 0 ÷ 1 | 0.000 | 83.2258 |
| $\Omega_M$ | Yes | 0 ÷ 1 | 1.000 | |
| $\Omega_K$ | No | 0 | 0 | |

It can be seen from the table that, assuming the evolution of SNe1a, the best fit (the smallest Chi$^2$) of the flat universe model $\Lambda$CDM with observational data is obtained at $\Omega_\Lambda=0$. A comparison of Chi$^2$ in Tables 2 and 3 shows that its value is smaller in Table 3, i.e., assuming the evolution of the absolute magnitude of SNe1a, we obtain a better fit between theory and observation. In this case, we need a change in the absolute magnitude of SNe1a of only 0.3$^m$ for the time of the corresponding z=1 (approximately 5.3 Gyr). This value is consistent with the value obtained in Kang et al. (2020) (see Table 1).

Let's do an absolute magnitude test. The dependence of the absolute magnitudes of supernovae on the redshift at the values of the parameters $\varepsilon_c=0.304$, $\Omega_\Lambda=0.000$ and $\Omega_M=1.000$ is shown in Fig. 4.

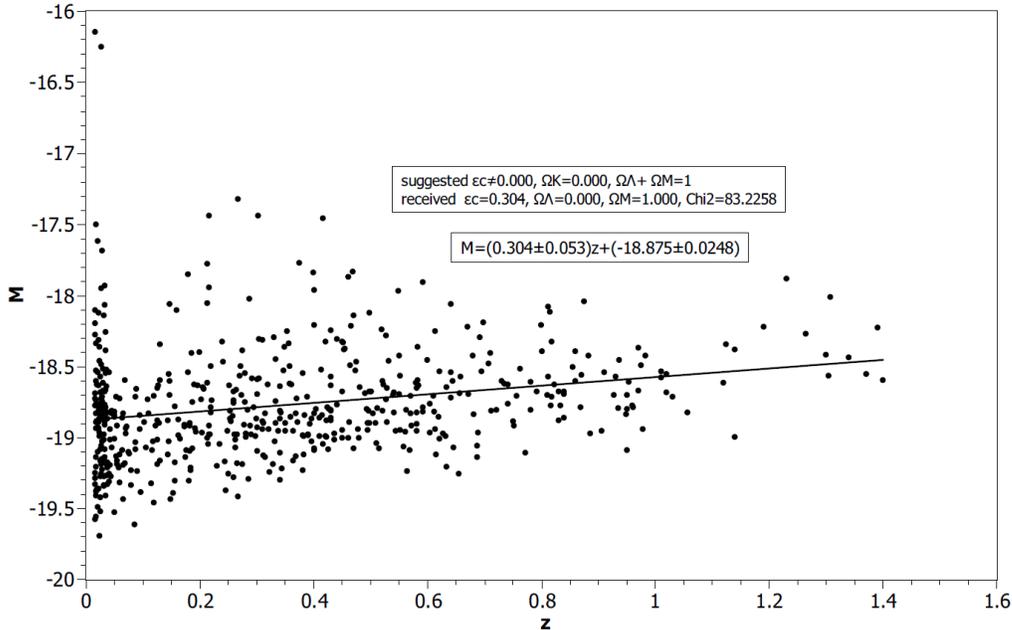

Fig. 4. Dependence of the absolute magnitudes of supernovae on the redshift at the values of the parameters $\varepsilon_c=0.304$, $\Omega_\Lambda=0.000$ and $\Omega_M=1.000$.

The figure shows that the dependence of the absolute magnitudes of SNe1a repeats what is assumed in advance (there is an evolution of the absolute magnitude). The slope of the dependence is the same as in the simulation.

5.3. Case $\Omega_\Lambda+\Omega_M+\Omega_K=1$, $\varepsilon_c=0$, estimated $M_0$, $\Omega_\Lambda$, $\Omega_M$, $\Omega_K$

Table 4 shows the result of searching for the values of the parameters $M_0$, $\Omega_\Lambda$, $\Omega_M$, $\Omega_K$ without the assumption that the universe is flat ($\Omega_\Lambda+\Omega_M+\Omega_K=1$) and without taking into account the evolution of the absolute magnitude SNe1a.

| Parameter | Variable | Search range | Parameter search result | Chi² |
|---|---|---|---|---|
| $M_0$ | Yes | -19.5 ÷ -18.0 | -18.881 | |
| $\varepsilon_c$ | No | 0 | 0 | |
| $\Omega_\Lambda$ | Yes | 0 ÷ 1 | 0.000 | 83.2808 |
| $\Omega_M$ | Yes | 0 ÷ 1 | 0.368 | |
| $\Omega_K$ | Yes | 0 ÷ 1 | 0.632 | |

As can be seen from the table, when in the general case ($\Omega_\Lambda+\Omega_M+\Omega_K=1$) we do not take into account the evolution of the absolute magnitude of SNe1a, the fraction of repulsive energy is 0, the fraction of gravitational material is approximately 0.37, which is consistent with the popular opinion about the fractions of dark and visible matter. The curvature of space is negative.

A check of the absolute magnitude test is shown in Fig. 5.

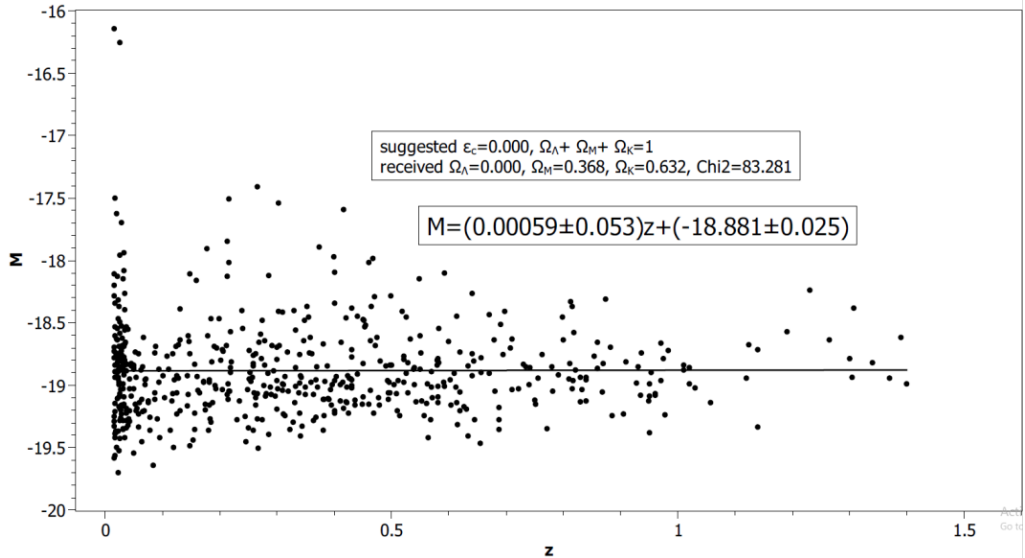

Fig.5. Absolute magnitude test for the case $\Omega_\Lambda+\Omega_M+\Omega_K=1$, $\varepsilon_c=0$

As can be seen in this case, the original assumption about the independence of the absolute magnitudes of the redshift is not violated.

5.4. Case $\Omega_\Lambda+\Omega_M+\Omega_K=1$, estimated $M_0$, $\varepsilon_c$, $\Omega_\Lambda$, $\Omega_M$, $\Omega_K$

In Table 5. the result of the search for the values of the parameters $M_0$, $\Omega_\Lambda$, $\Omega_M$, $\Omega_K$ without restrictions on the curvature of the universe ($\Omega_\Lambda+\Omega_M+\Omega_K=1$) is given with the assumption of the evolution of the absolute magnitude SNe1a.

Table 5. The result of searching for the values of the parameters $M_0$, $\Omega_\Lambda$, $\Omega_M$, $\Omega_K$ without restriction on the curvature of the universe ($\Omega_\Lambda+\Omega_M+\Omega_K=1$) with the assumption of the evolution of the absolute magnitude SNe1a.

| Parameter | Variable | Search range | Parameter search result | Chi² |
|---|---|---|---|---|
| $M_0$ | Yes | -19.5 ÷ -18.0 | -18.875 | |
| $\varepsilon_c$ | Yes | -1 ÷ 1 | 0.304 | |
| $\Omega_\Lambda$ | Yes | 0 ÷ 1 | 0.000 | 83.2258 |
| $\Omega_M$ | Yes | 0 ÷ 1 | 1.000 | |
| $\Omega_K$ | Yes | 0 ÷ 1 | 0.000 | |

As can be seen from the table, the simulation gives the same result as Case $\Omega_K=0$, $\Omega_\Lambda+\Omega_M=1$ (see 5.2.). That is, without initially assuming that the universe is flat, the largest probable estimate of cosmological parameters is obtained precisely with a flat universe.

As for the absolute magnitude test, it coincides with the graph in Fig. 4.

Thus, we can say that, under the assumption of the evolution of type 1a supernovae, the ΛCDM model gives two important results:
a - the universe is flat
b - only gravitational material is present in it – the fraction of dark energy is equal to zero. The universe is expanding at a slower pace.

5.5. Case Λ=0, $\varepsilon_c$=0, estimated $M_0$, $q_0$

In Table 6, the result of the search for the values of the parameters $M_0$, $q_0$ for the model with a zero cosmological constant (Λ = 0) is given without the assumption of the evolution of the absolute magnitude SNe1a.

Table 6. The result of the search for the best solutions for the model with a zero cosmological constant (Λ=0) without assuming the evolution of the absolute magnitude of SNe1a.

| Parameter | Variable | Search range | Parameter search result | Chi² |
|---|---|---|---|---|
| $M_0$ | Yes | -19.5 ÷ -18.0 | -18.881 | |
| $\varepsilon_c$ | No | 0 | 0 | 83.2808 |
| $q_0$ | Yes | 0 ÷ 0.5 | 0.184 | |

In essence, this is similar to the case in 5.3. Therein, space has a negative curvature and contains only gravitational matter (the sum of visible and invisible matter is 0.37, which is consistent with many other studies).

Fig. 6 shows the absolute magnitude test. As can be seen from the figure, with the results obtained ($M_0$=-18.881, $q_0$=0.184), there is no dependence of the absolute magnitude of SNe1a on z; this repeats the original assumption of independence between them.

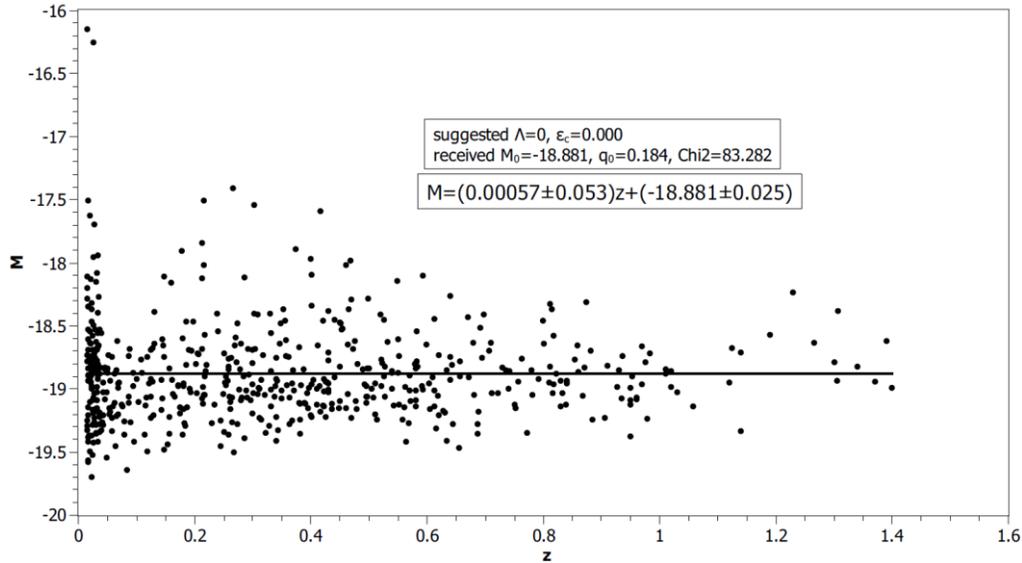

Fig.6. Dependence of the absolute magnitude of SNe1a on z at $M_0= -18.881$, $q_0=0.184$ for a universe with a zero cosmological constant.

5.6. Case $\Lambda=0$, $M_0$, $\varepsilon_c$, $q_0$ are estimated

Table 7 shows the search result for the values of the parameters $M_0$, $q_0$ and $\varepsilon_c$ for the model with zero cosmological constant ($\Lambda=0$) with the assumption of the evolution of the absolute magnitude SNe1a.

The absolute magnitude test, i.e., the dependence of the absolute magnitude of SNe1a on the redshift at the obtained values of the parameters $\varepsilon_c=0.304$, $q_0=0.5$, is shown in Fig. 7.

As can be seen from Fig. 7 the absolute magnitude of supernovae SNe1a depends on the distance as obtained from the simulation, which repeats the original assumption.

In fact, we get an analogy of the cases in 5.2 and 5.4. We obtain that, under the assumption of the evolution of magnitude of supernova 1a, both hypotheses give the same result, which consists of the fact that the universe is flat and consists only of gravitational material.

Table 7. The result of the search for the values of the parameters $M_0$, $q_0$ and $\varepsilon_c$ for the model with a zero cosmological constant ($\Lambda = 0$) with the assumption of the evolution of the absolute magnitude of SNe1a.

| Parameter | Variable | Search range | Parameter search result | Chi² |
|---|---|---|---|---|
| $M_0$ | Yes | -19.5 ÷ -18.0 | -18.875 | |
| $\varepsilon_c$ | Yes | -1 ÷ 1 | 0.304 | 83.2258 |
| $q_0$ | Yes | 0 ÷ 0.5 | 0.500 | |

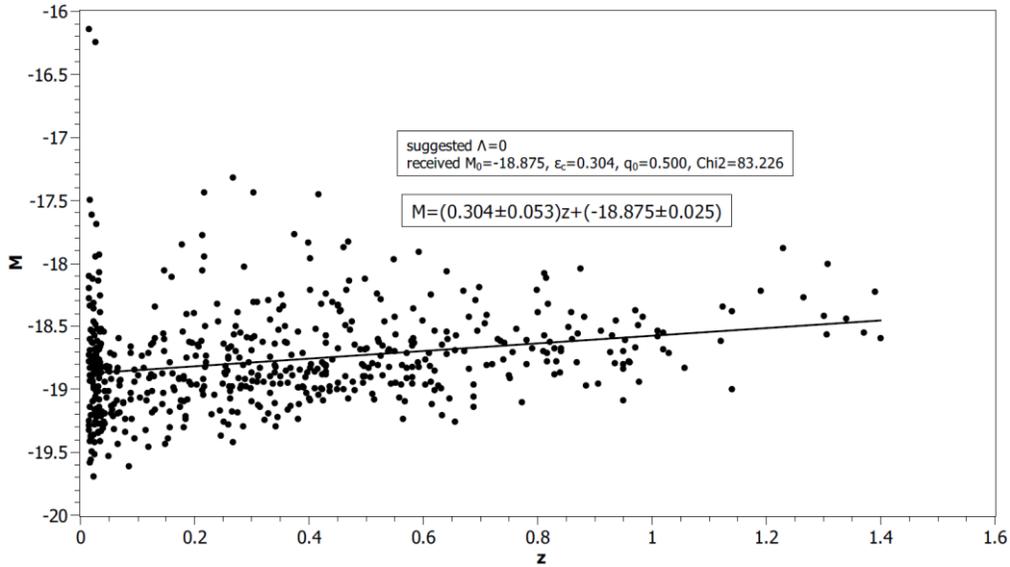

Fig.7. Dependence of the absolute magnitude of SNe1a on the redshift at the obtained values of the parameters $\varepsilon_c=0.304$, $q_0=0.5$

5.7. Verification of the evolution of SNe1a at different z.

It is interesting to check the evolution of SNe1a at different z. To do this, we divided the Union 2 sample into two parts – a subsample with a redshift up to z=0.5 and a subsample with z≥0.5.

5.7.1. The case $\Lambda=0$, $z=0.00 \div 0.5$, N=403 and the case $\Lambda=0$, $z=0.5 \div 1.5$, N=154 are evaluated $M_0$, $\varepsilon_c$, $q_0$.

In Table 8. the result of the search for the values of the parameters $M_0$, $q_0$ and $\varepsilon_c$ for a model with a zero cosmological constant ($\Lambda=0$) is given with the assumption of the evolution of the absolute magnitude of SNe1a for "nearby" stars (z=0.00 ÷ 0.5). In Table. 9. the same is shown for distant stars (z=0.5÷1.5).

It can be seen from the tables that the smallest $Chi^2$ is obtained at $\varepsilon_c=0.399$ and $\varepsilon_c=0.403$, respectively. In these cases, we get the value 0.5 for $q_0$, i.e., a flat universe.

Table 8. The result of the search for the values of the parameters, $q_0$ and $\varepsilon_c$ for the model with zero cosmological constant ($\Lambda=0$) with the assumption of the evolution of the absolute magnitude of SNe1a for "nearby" stars (z=0.00 ÷ 0.5).

| Parameter | Variable | Search range | Parameter search result | $Chi^2$ |
|---|---|---|---|---|
| $M_0$ | Yes | -19.5 ÷ -18.0 | -18.886 | |
| $\varepsilon_c$ | Yes | -1 ÷ 1 | 0.399 | 72.2283 |
| $q_0$ | Yes | 0 ÷ 0.5 | 0.500 | |

Table 9. The same as in Table 8, for distant stars (z=0.5 ÷ 1.5).

| Parameter | Variable | Search range | Parameter search result | $Chi^2$ |
|---|---|---|---|---|
| $M_0$ | Yes | -19.5 ÷ -18.0 | -18.970 | |
| $\varepsilon_c$ | Yes | -1 ÷ 1 | 0.403 | 10.7607 |
| $q_0$ | Yes | 0 ÷ 0.5 | 0.500 | |

5.7.2. The case $\Omega_\Lambda+\Omega_M+\Omega_K=1$, z=0.00 ÷ 0.5, N=403 and the case $\Omega_\Lambda+\Omega_M+\Omega_K=1$, z=0.50 ÷ 1.50, N=154. Are evaluated $M_0$, $\varepsilon_c$, $\Omega_\Lambda$, $\Omega_M$, $\Omega_K$

In Table 10. the result of the search for the values of the parameters $M_0$, $\Omega_\Lambda$, $\Omega_M$, $\Omega_K$ and $\varepsilon_c$ for the universe without space curvature restrictions ($\Omega_\Lambda+\Omega_M+\Omega_K=1$) is given with the assumption of the evolution of the absolute magnitude of SNeIa for "nearby" stars (z=0.00 ÷ 0.5). In Table 11 the same is given for distant stars (z=0.5 ÷ 1.5).

Table 10. The result of the search for the values of the parameters $M_0$, $\Omega_\Lambda$, $\Omega_M$, $\Omega_K$ and $\varepsilon_c$ for the universe without space curvature restrictions ($\Omega_\Lambda+\Omega_M+\Omega_K=1$) with the assumption of the evolution of the absolute magnitude of SNeIa for "nearby" stars (z=0.00 ÷ 0.5).

| Parameter | Variable | Search range | Parameter search result | Chi$^2$ |
|---|---|---|---|---|
| $M_0$ | Yes | -19.5 ÷ -18.0 | -18.886 | 72.2283 |
| $\varepsilon_c$ | Yes | -1 ÷ 1 | 0.399 | |
| $\Omega_\Lambda$ | Yes | 0 ÷ 1 | 0.000 | |
| $\Omega_M$ | Yes | 0 ÷ 1 | 1.000 | |
| $\Omega_K$ | No | 0 | 0 | |

Table 11. The same as in Table 10 for distant stars (z=0.5 ÷ 1.5).

| Parameter | Variable | Search range | Parameter search result | Chi$^2$ |
|---|---|---|---|---|
| $M_0$ | Yes | -19.5 ÷ -18.0 | -18.970 | 10.7607 |
| $\varepsilon_c$ | Yes | -1 ÷ 1 | 0.403 | |
| $\Omega_\Lambda$ | Yes | 0 ÷ 1 | 0.000 | |
| $\Omega_M$ | Yes | 0 ÷ 1 | 1.000 | |
| $\Omega_K$ | Yes | 0 ÷ 1 | 0 | |

As can be seen from Tables 8-11, the simulation shows that the evolution of SNe1a is observed for both nearby and distant supernovae. The direction of evolution is also the same for nearby and distant supernovae - young supernovae are dimmer.

6. Significant influence of the absolute magnitude of type 1a supernovae on cosmological parameters.

In this section, using the created computer model, we also study the influence of the absolute magnitude of supernovae 1a on the obtained cosmological parameters. It turned out that the cosmological parameters are very sensitive to even a small change in this value.

This issue is very important because, when estimating cosmological parameters, researchers use the absolute magnitude of supernovae obtained by only a few stars. As we noted above, the distribution of the absolute magnitude of type 1a supernovae is very wide, and at first glance it can be seen (you can compare figures 2 and 3) that the value of the cosmological parameters strongly depends on the absolute magnitude of supernovae. Let's try to study this issue in more detail.

Figure 8 shows a graph of dependence $\Omega_\Lambda$, $\Omega_M$ on M. This graph is constructed for the ΛCDM model for a flat universe ($\Omega_\Lambda+\Omega_M=1$) and no evolution ($\varepsilon_c=0$). The graph shows the values of M corresponding to three combinations of cosmological parameters:
a) $\Omega_\Lambda = 0.7$, $\Omega_M = 0.3$ obtained at M = -19.11;
b) $\Omega_\Lambda = 0$, $\Omega_M = 1$ obtained at M = -18.71;
c) $\Omega_\Lambda = 0.397$, $\Omega_M = 0.603$ obtained at M = -18.90;
At the same time, as shown above, the best solution for a flat universe, without taking into account evolution, was obtained in the latter case (see Table 2).

The difference in the absolute magnitude of supernovae 1a with combinations of a and b is: 19.11-18.71 = 0.4 magnitudes, while the standard deviation of the distribution of absolute magnitudes of SNe1a is 0.7 magnitudes. At the same time, the difference in magnitude between the combinations a and c is only 0.2.

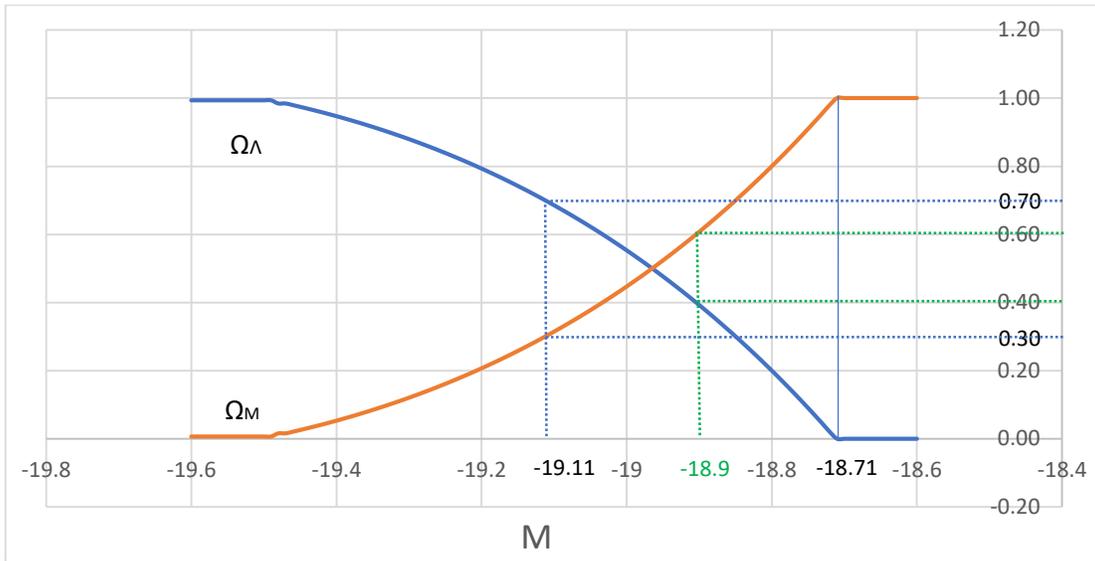

Figure 8. Plot, $\Omega_\Lambda$, $\Omega_M$ versus M calculated for the $\Lambda$CDM model for a flat universe ($\Omega_\Lambda+\Omega_M=1$). As can be seen from the figure, a change in M by only $0.4^m$ (from -19.11 to -18.71) leads to a change in parameters from $\Omega_\Lambda=0.7$ and $\Omega_M=0.3$ to $\Omega_\Lambda=0$ and $\Omega_M=1$

Tab. 12. Excerpt from the table from the simulation

| M | $\Omega_\Lambda$ | $\Omega_M$ | Chi² |
|---|---|---|---|
| -18.9 | 0.39 | 0.61 | 83.74 |
| -19.03 | 0.60 | 0.40 | 86.85 |
| -19.04 | 0.61 | 0.39 | 87.36 |
| -19.05 | 0.62 | 0.38 | 87.92 |
| -19.06 | 0.64 | 0.36 | 88.53 |
| -19.07 | 0.65 | 0.35 | 89.17 |
| -19.08 | 0.66 | 0.34 | 89.86 |
| -19.09 | 0.67 | 0.33 | 90.60 |
| -19.1 | 0.69 | 0.31 | 91.38 |
| -19.11 | 0.70 | 0.30 | 92.20 |
| -19.12 | 0.71 | 0.29 | 93.07 |
| -19.13 | 0.72 | 0.28 | 93.98 |
| -19.14 | 0.73 | 0.27 | 94.95 |
| -19.15 | 0.74 | 0.26 | 95.95 |
| -19.16 | 0.75 | 0.25 | 97.01 |
| -19.17 | 0.76 | 0.24 | 98.11 |
| -19.18 | 0.77 | 0.23 | 99.26 |
| -19.19 | 0.78 | 0.22 | 100.46 |
| -19.2 | 0.79 | 0.21 | 101.71 |
| -19.21 | 0.80 | 0.20 | 103.01 |

Table 12 shows an excerpt from the table from the simulation. The first line shows the case of the best fit between theory and observation. The rest of the lines show the results in the vicinity of the point $\Omega_\Lambda$=0.7 and $\Omega_M$=0.3. It can be seen that when M changes by only 0.1 in one direction or another, we obtain the value of $\Omega_\Lambda$ from 0.6 to 0.8. Such a difference in the values of $\Omega_\Lambda$ significantly changes the idea of cosmology. The change in M by 0.1 is well below the range of this value used in various papers.

Thus, the dependence of the values of the parameters $\Omega_\Lambda$ and $\Omega_M$ on the accepted absolute magnitude M SNe1a is very strong, and therefore we must be extremely careful when determining M. According to the authors of this article, the determination of the absolute magnitude of supernovae should be the subject of a simulation using the entire sample of type 1a supernovae. As we saw above, this approach does not violate the original assumption that the absolute magnitudes of supernovae depend on the redshift and, therefore, gives the correct value of the parameters $\Omega_\Lambda$ and $\Omega_M$.

7. Conclusion

In the previous article, we studied the value of cosmological parameters in the ΛCDM and CDM models with the assumption that the absolute magnitude of type 1a supernovae is independent of distance. The values of the cosmological parameters were estimated on the basis of the Hubble diagram. The values of these parameters were determined for which the best fit between the theoretical curve of the Hubble diagram and observational data was obtained. Pearson's goodness-of-fit test or Chi$^2$ (Chi-square) test was used.

In this article, we study the case with the assumption of the evolution of the absolute magnitude. We accept that the dependence of the absolute magnitude on the redshift is linear.

It turns out that when the evolution of the absolute magnitudes of supernovae is taken into account, a better fit between theory and observation is obtained.

The main difference between the approaches in our works and the works of other authors is that we estimate the average absolute magnitude of supernovae in the course of simulation, while the authors of other works take into account the average absolute magnitude of these stars, previously obtained by several well-studied stars. We pay attention to the fact that the distribution of the absolute magnitude of supernovae 1a is very wide (Ashall et al. 2016). This makes it incorrect to use the latter approach. In addition, as the simulation shows, the result strongly depends on the assumed absolute magnitude of the stars.

In the case of a flat universe ($\Omega_M$+$\Omega_\Lambda$=1), the best fit between theory and observation is given by the value of the evolution coefficient $\varepsilon_c$=0.304. In this case, for the cosmological parameters we obtain $\Omega_\Lambda$=0.000, $\Omega_M$=1.000. And for the absolute magnitude of supernovae 1a, -18.875 was obtained. Naturally, this result exactly matches the simulation result for the model with a zero cosmological constant ($\varepsilon_c$=0.304, $q_0$= 0.500, $M_0$=-18.875).

The Friedmann-Robertson-Walker model is also studied without restriction on space curvature ($\Omega_M$+$\Omega_\Lambda$+$\Omega_K$=1). Within the framework of this model, we obtain the following values $\varepsilon_c$=0.304, $\Omega_\Lambda$=0, $\Omega_M$=1.000, $\Omega_K$=0.000, $M_0$=-18.875. Those, the general case also leads to a flat Universe model ($\Omega_K$=0.000).

In the framework of this work, the degree of influence of the absolute magnitude M of supernovae of type 1a on the cosmological parameters is also investigated. In particular, it was found that a change in this value by only 0.4$^m$ (from -19.11 to -18.71) leads to a change in the parameters from $\Omega_\Lambda$=0.7 and $\Omega_M$=0.3 to $\Omega_\Lambda$=0 and $\Omega_M$=1. As we noted above, the distribution of the absolute magnitudes of supernovae 1a has a rather large width, which leads us to think that we must be very careful when accepting the absolute magnitude of supernovae. We have come to the conclusion that the absolute magnitude of supernovae must also be subject to simulation. This is the main reason that leads to a discrepancy between our results and the results obtained by other authors, who took as a basis the value of the absolute magnitude of supernovae calculated from a small number of stars.

The validity of our results is substantiated by the absolute magnitude test we proposed in the previous article (Paper I). In essence, this is a test proving that the simulation does not violate the initially accepted dependence of the absolute magnitudes of supernovae on the redshift.

The main results of this work are the following:
a. Under the assumption of the evolution of supernovae SNe1a, the ΛCDM model describes the observational data better than under the assumption that the absolute magnitudes of SNe1a are independent of redshift. In this case, a small evolution is obtained (ΔM=0.304 during the time of the corresponding z=1). Young supernovae are dimmer. Evolution is observed for both nearby and distant stars.
    b. The universe turns out to be flat, even if this constraint is not initially introduced.
    c. There is only gravitational matter in the universe.
    d. The expansion of the universe is slowing down.

    The main difference between our and other authors' approaches is that we pay attention to two facts in the nature of SNe1a. The first is that there is a critical dependence of the values of cosmological parameters obtained during the simulation on the absolute magnitude of SNe1a. The second is that the distribution of absolute magnitudes of supernovae is very wide.
    These facts lead us to the conclusion that the found cosmological parameters based on the absolute magnitude of supernovae 1a predetermined from several stars are not accurate and that it is necessary to find the cosmological parameters and absolute magnitude of supernovae simultaneously in the simulation process using the full sample of supernovae 1a.